\documentclass[prb,twocolumn,citeautoscript]{revtex4-2}

\usepackage{grffile}
\usepackage{graphicx}
\usepackage{amsmath}
\usepackage{amssymb}
\usepackage{color}
\usepackage{mathrsfs}
\usepackage{braket}
\usepackage{textcomp}
\usepackage{hyperref}

\usepackage{bm}

\begin{document}

\title{Sine-Gordon dynamics in spin transport}
\author{A.-M. Visuri}
\email{avisuri@uni-bonn.de}
\affiliation{Physikalisches Institut, University of Bonn, Nussallee 12, 53115 Bonn, Germany}

\begin{abstract}

We study spin transport in a one-dimensional finite-length wire connected to fermionic leads. The interacting wire is described by the sine-Gordon model while the leads are either noninteracting or interacting Luttinger liquids. We calculate the spin current driven by a spin bias in the nonlinear regime by solving numerically the classical equation of motion, and we find that the cosine term in the sine-Gordon model gives rise to an oscillating spin current when the spin bias exceeds its critical value. We discuss the results in connection with transport experiments with ultracold atoms.

\end{abstract}

\maketitle

\section{Introduction}

The sine-Gordon model is a relativistic field theory ubiquitous in physics, applied both in high-energy and condensed-matter physics, especially in the description of low-dimensional systems~\cite{EsslerKonik2005, GogolinTsvelik2004, Giamarchi_one_dimension2003}. In the condensed-matter context, the model provides the low-energy description of systems as varied as spin chains~\cite{HuddartLancaster2021}, long Josephson junctions~\cite{Lomdahl1985, Cuevas-MaraverFloyd2014, DeSantisValenti2022} and Josephson-junction arrays~\cite{RoySaleur2019, RoySaleur2021}, fermions and bosons in one-dimensional periodic potentials~\cite{HallerNagerl2010, Lebrat_band_and_correlated2018} and coupled quasi-one-dimensional Bose-Einstein condensates~\cite{HofferberthSchmiedmayer2007, SchweiglerSchmiedmayer2017, PigneurSchmiedmayer2018, ZacheSchmiedmayer2020, SchweiglerSchmiedmayer2021, JiSchmiedmayer2022}. In one dimension, the spin and charge degrees of freedom of fermions decouple in the low-energy limit, and the collective spin degree of freedom is generally described by the sine-Gordon model~\cite{Giamarchi_one_dimension2003}. 

The model has been extensively studied both numerically and analytically. Due to its integrability, many quantities can be solved exactly~\cite{LukyanovZamolodchikov1997}. The excitation spectrum has been mapped in terms of solitons, antisolitons, and their bound states called breathers~\cite{Rajaraman1982}.
An experiment with two coupled quasi-one-dimensional Bose-Einstein condensates, where the relative phase realizes the sine-Gordon model~\cite{HofferberthSchmiedmayer2007, SchweiglerSchmiedmayer2017, PigneurSchmiedmayer2018, ZacheSchmiedmayer2020, SchweiglerSchmiedmayer2021, JiSchmiedmayer2022}, has inspired extensive theoretical work. Its ground states~\cite{KasperDemler2020} as well as dynamics due to periodic driving~\cite{GritsevDemler2007, LovasOrignac2022} and quenches~\cite{DallaTorrePolkovnikov2013, FoiniGiamarchi2015, FoiniGiamarchi2017, KukuljanTakacs2018, vanNieuwkerkEssler2020, RuggieroGiamarchi2021, HorvathTakacs2022, ChelpanovaMarino2023, SzaszSchagrinTakacs2024} have been described using a wide range of methods based on semiclassical approximations, Hilbert-space truncation, or matrix product states. Similar realizations have been proposed with coupled spin chains~\cite{WyboBastianello2022} and one-dimensional Bose-Hubbard chains~\cite{WyboKnap2023}. 
Generalized hydrodynamics descriptions have recently illuminated large-scale fluctuations~\cite{DelVecchioBastianello2023}, thermodynamics, and the transport of solitons and antisolitons, in particular the spreading of the topological charge in a bipartitioning protocol~\cite{BertiniKormos2019, KochBastianello2023, NagyTakacs2023, Bastianello2023}. Experimentally, the dynamics of soliton and antisoliton excitations have been observed in spin chain materials by high-field spectroscopy~\cite{Zvyagin2021}.

Another important nonequilibrium situation arises in a two-terminal transport measurement, commonly used to characterize the properties of solid-state systems. 
Transport measurements reveal the insulating, conducting, or superconducting properties of materials, essential for the design of any device. 
To measure steady-state charge transport, the system is coupled to leads, or reservoirs, at different chemical potentials, which induces an electric current. While a chemical potential bias leads to charge transport, a temperature bias drives a heat current and a spin bias drives a spin current. 
Two-terminal transport measurements have recently also been realized in experiments with ultracold atoms, where two atom cloud reservoirs are coupled by a point contact or a narrow channel~\cite{KrinnerBrantut2017}. This setup was used for example to measure particle and spin conductances~\cite{Krinner_spin_and_particle2016} and to study transport through a short one-dimensional periodic potential~\cite{Lebrat_band_and_correlated2018},
where the charge sector is described by the sine-Gordon model in the low-energy limit. 

From a theoretical point of view, the Landauer-B\"uttiker formalism underpins our understanding of quantum coherent transport in noninteracting mesoscopic systems~\cite{Landauer_spatial_variation1957, Landauer_electrical_resistance1970, Buttiker_multiprobe_conductors1988}.
It has been generalized to interacting particles using nonequilibrium Green's functions~\cite{Meir_interacting1992}.
In the paradigmatic example of an interacting one-dimensional Tomonaga-Luttinger liquid (TLL) wire coupled to leads, the conductance can be solved analytically using a bosonized description~\cite{Safi_transport1995, Maslov_conductance1995, Ponomarenko_conductance1995}. 
Such a system with noninteracting leads is known to have a conductance equal to the conductance quantum independent of the interactions in the TLL wire.
When the leads are interacting TLLs, the conductance is
proportional to the Luttinger parameter of the leads.
While for spinless fermions, the low-energy limit maps to a TLL, for fermions with spin, the backscattering of opposite-spin fermions gives rise to the sine-Gordon model
even in the absence of any external potentials. 
An interesting question therefore is how spin is transported through an interacting wire of fermions with spin.

Spin transport in a sine-Gordon wire coupled to leads was previously analyzed via renormalization group in the linear-response regime~\cite{Ponomarenko_spin_transport_Mott2014, VisuriGiamarchi2020}. 
Here, we focus instead on the real-time dynamics, characterized by the spin density and spin current, in the nonlinear regime at finite spin bias. We adopt an approach similar to that of Refs.~\cite{Maslov_conductance1995, Lebrat_band_and_correlated2018}.
Namely, the spin degree of freedom is driven by a magnetic-field gradient, and we compute the spin current by solving numerically the classical equation of motion of the corresponding forced sine-Gordon model. 
We find that the spin current either is damped to zero or oscillates with a nonzero average value, depending on the magnitude of the magnetic-field gradient with respect to the coefficient of the cosine term in the sine-Gordon model. We characterize the oscillation frequency and amplitude and compute the differential conductance of the interacting wire both for noninteracting leads and leads that are interacting Luttinger liquids.

The paper is organized as follows: We introduce the model of the interacting sine-Gordon wire, the observables used to characterize spin transport, and the classical equation of motion in Sec.~\ref{sec:model}. Section~\ref{sec:general_solution} presents the solution of the equation of motion for generic parameters, while Sec.~\ref{sec:experimental} discusses its possible connection to the cold-atom transport experiment of Refs.~\cite{Krinner_spin_and_particle2016, Lebrat_band_and_correlated2018}. Finally, the discussion is expanded in Sec.~\ref{sec:discussion}, conclusions are presented in Sec.~\ref{sec:conclusions}, and technical details are found in the Appendices.

\section{Model}
\label{sec:model}

\begin{figure}
\centering
\includegraphics[width=0.72\linewidth]{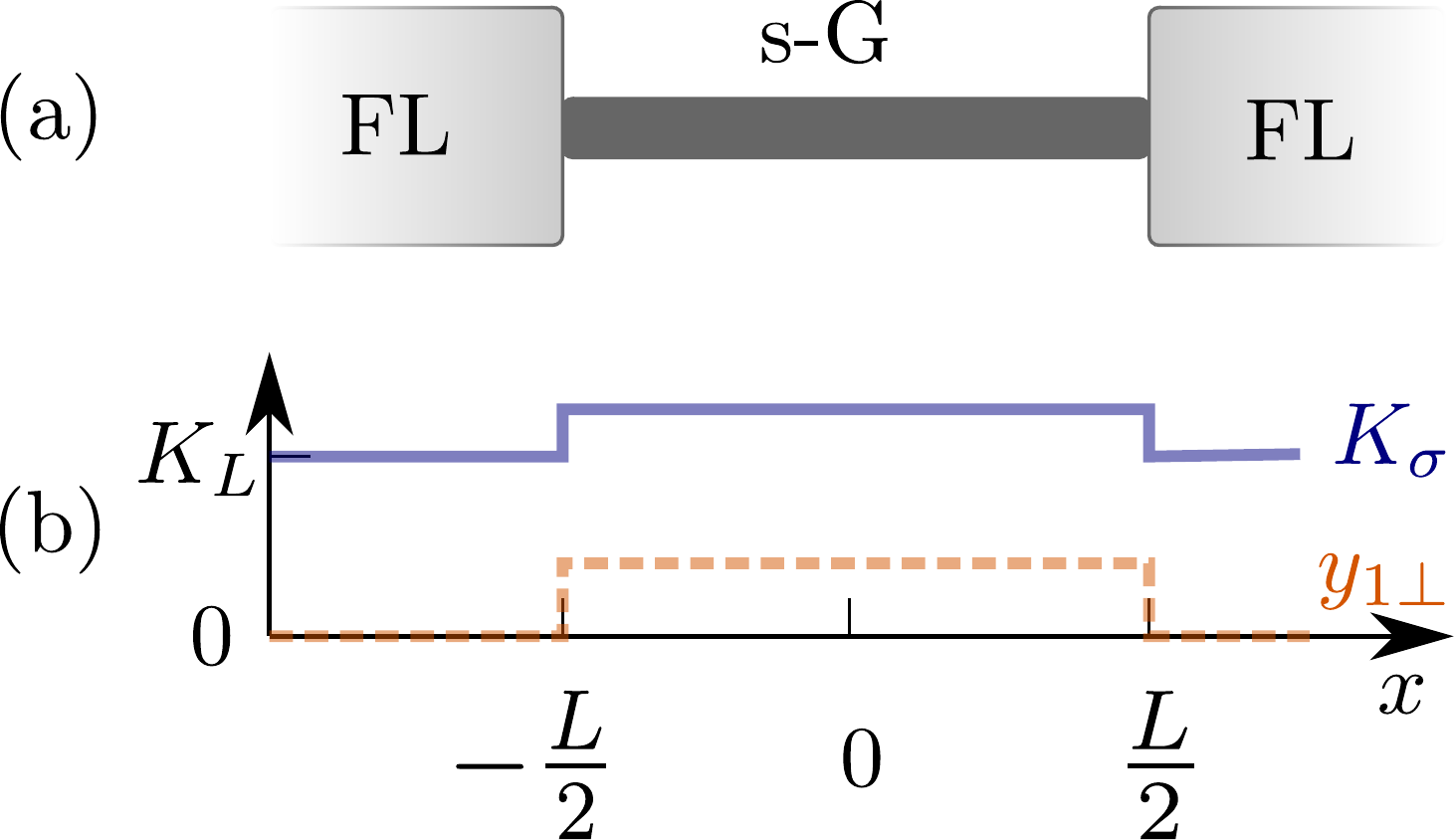}
\caption{(a) A one-dimensional wire of length $L$ is coupled to Fermi-liquid (FL) leads, modeled by noninteracting 1D systems. The interacting wire is described by the sine-Gordon (s-G) model [Eqs.~\eqref{eq:quadratic_Hamiltonian} and~\eqref{eq:sine-Gordon}], and the noninteracting leads are described by the quadratic Hamiltonian~\eqref{eq:quadratic_Hamiltonian}. (b) The Luttinger parameter is $K_{\sigma}$ in the wire and $K_{\sigma} = K_L$ in the leads, while the coupling $y_{1\perp}$ is zero in the leads and nonzero in the wire. To model FL leads, we set $K_L = 1$. Modified from Ref.~\cite{VisuriGiamarchi2020}.}
\label{fig:geometry}
\end{figure}

We consider an interacting wire of finite size $L$, connected to infinite leads on either side, as depicted in Fig.~\ref{fig:geometry}(a). We focus on the case where the leads are in the Fermi-liquid state and are modeled as one-dimensional noninteracting systems with Luttinger parameter ${K_L = 1}$. 
In Sec.~\ref{sec:different_K_v}, we also comment briefly on the case of interacting TLL leads, for which the description is identical apart from $K_L \neq 1$.
We use a bosonized description where the interacting and noninteracting regions are taken into account by varying the relevant parameters [Fig.~\ref{fig:geometry}(b)].
The Hamiltonian for the interacting wire is of the sine-Gordon form. This model describes the low-energy properties of a wide class of microscopic models in one dimension, and for the most part, we do not limit the discussion to a specific microscopic model.
In Sec.~\ref{sec:experimental}, however, we discuss the parameter regime relevant for the experiment of Ref.~\cite{Krinner_spin_and_particle2016}, and in this context, consider the one-dimensional continuum Hamiltonian
\begin{align}
\begin{split}
H = -\frac{\hbar^2}{2m} &\sum_{s = \uparrow, \downarrow} \int dx \psi_s^{\dagger}(x) \frac{\partial^2}{\partial x^2} \psi_s^{\phantom{\dagger}}(x) \\
&+ g_{1\perp} \int dx \psi^{\dagger}_{\downarrow}(x) \psi^{\dagger}_{\uparrow}(x) 
\psi^{\phantom{\dagger}}_{\uparrow}(x) \psi^{\phantom{\dagger}}_{\downarrow}(x)
\end{split}
\label{eq:continuum_hamiltonian}
\end{align}
for fermionic atoms with contact interactions. Here, $\hbar$ is the reduced Planck constant, $m$ the atom mass, $g_{1\perp}$ the coupling constant for the backscattering of fermions with opposite spin, and $\psi_s^{\dagger}$ ($\psi_s^{\phantom{\dagger}}$) the fermionic field operator that creates (destroys) a particle with spin $s = \uparrow, \downarrow$. We adopt natural units where $\hbar = k_B = 1$.

\subsection{Bosonization}

In the low-energy limit, Eq.~\eqref{eq:continuum_hamiltonian} maps onto the effective bosonized model $H = H_{\rho}^0 + H^0_{\sigma} + H_{\sigma}'$. Here, $\rho$ and $\sigma$ denote the collective charge and spin degrees of freedom, respectively. In the absence of external potentials, charge and spin are decoupled and the term $H_{\nu}^0$ with $\nu = \rho, \sigma$ is the TLL Hamiltonian
\begin{equation}
H_{\nu}^0 = \frac{1}{2\pi} \int_{-\infty}^{\infty} dx \left[ v_{\nu} K_{\nu}\left( \partial_x \theta_{\nu} \right)^2 + \frac{v_{\nu}}{K_{\nu}}\left( \partial_x \phi_{\nu} \right)^2 \right].
\label{eq:quadratic_Hamiltonian}
\end{equation}
This model is quadratic in the boson fields $\phi_{\nu}(x, t)$ and $\theta_{\nu}(x, t)$. The collective charge and spin excitations have the velocities $v_{\rho}$ and $v_{\sigma}$, which are in general different from each other. The TLL represents a critical system where correlation functions decay as power laws, and the Luttinger parameters $K_{\rho}$ and $K_{\sigma}$ appear in the exponents. The velocities and Luttinger parameters are determined by the parameters, such as interactions, of the original microscopic model. 

While the charge excitations are described by the quadratic TLL model, the spin Hamiltonian has an additional term $H_{\sigma}'$ arising from the backscattering of fermions with opposite spin,
\begin{equation}
H_{\sigma}' = \frac{2 g_{1\perp}}{(2\pi\alpha)^2} \int_{-\frac{L}{2}}^{\frac{L}{2}} dx \cos \left( \phi_{\sigma} \right).
\label{eq:sine-Gordon}
\end{equation}
Here, $g_{1\perp}$ is the coupling constant for backscattering and $\alpha$ is a short-distance cutoff. The Hamiltonian $H_{\sigma}^0 + H_{\sigma}'$ with the quadratic and cosine terms is known as the sine-Gordon Hamiltonian~\cite{Giamarchi_one_dimension2003, Cuevas-MaraverFloyd2014}. 
Additionally, we include a magnetic field which is different in the two reservoirs, so that there is a magnetic-field gradient along the wire. This is analogous to the electric field considered in Refs.~\cite{Maslov_conductance1995, Lebrat_band_and_correlated2018}, and is taken into account by the force term 
\begin{equation}
H_{h} = \frac{1}{\sqrt{2}\pi} \int_{-\frac{L}{2}}^{\frac{L}{2}} dx h(x) \partial_x \phi_{\sigma}
\end{equation}
Here, $h = g \mu_B \mathcal{H}$, where $\mathcal{H}$ is the magnetic field, $g$ the Land\'e factor, and $\mu_B$ the Bohr magneton, so that $h$ has the dimension of energy. In the following, we will use the term ``magnetic field'' to refer to $h$. The magnetic field couples to the spin degree of freedom but not the charge. The total spin Hamiltonian is now the forced sine-Gordon model
\begin{equation}
H_{\sigma} = H_{\sigma}^0 + H_{\sigma}' + H_{h}.
\label{eq:total_hamiltonian}
\end{equation}
We model the Fermi-liquid leads by noninteracting one-dimensional regions with $K_{\rho} = K_{\sigma} = 1$. 
As we do not consider a chemical potential bias, there is no charge transport in the wire and we limit the discussion to spin transport. 

While the TLL is gapless, the cosine term in the sine-Gordon Hamiltonian can give rise to an energy gap. When the interactions are attractive, the spin gap can be understood as the pairing energy of fermions with opposite spin. In terms of the bosonized Hamiltonian, in the limit of strong coupling, the field $\phi_{\sigma}$ is fixed to one of the minima of the cosine.
An expansion around the minimum gives a quadratic mass term, where the mass is equal to the spin gap in this limit.
A gapped and a gapless phase therefore exist in the thermodynamic limit---the gapped phase is generally insulating while the gapless phase supports the ballistic transport of excitations. The transport properties can however vary when the system is of finite length~\cite{Ponomarenko_spin_transport_Mott2014, VisuriGiamarchi2020}. 
A periodic potential leads similarly to a sine-Gordon Hamiltonian for the charge degree of freedom, and the charge conductance in the presence of umklapp scattering from a periodic potential was considered previously in Refs.~\cite{Ponomarenko_treshold_features1997, Ponomarenko_Mott_insulator1998, Ponomarenko_spin_gap_insulators2000, Ponomarenko_spin_transport_Mott2014, Lebrat_band_and_correlated2018}.

\subsection{Observables}

To characterize spin transport in the wire, we compute the 
spin current $j_{\sigma} = j_{\uparrow} - j_{\downarrow}$ obtained from the continuity equation $\partial_t \sigma(x, t) + \partial_x j_{\sigma} = 0$. Here, $\sigma(x, t)$ is the spin density $\sigma = \rho_{\uparrow} - \rho_{\downarrow}$. We use the relations $\rho_{s} = -\partial_x \phi_s/\pi$, with $s = \uparrow, \downarrow$, and 
\begin{equation}
\phi_{\sigma} = \frac{1}{\sqrt{2}}(\phi_{\uparrow} - \phi_{\downarrow})
\end{equation}
to obtain the spin density as
\begin{equation}
\sigma(x, t) = -\frac{\sqrt{2}}{\pi} \partial_x \phi_{\sigma}(x, t).
\label{eq:spin_density}
\end{equation}
The spin current is written in terms of the time derivative of the field $\phi_{\sigma}$,
\begin{equation}
j_{\sigma}(x, t) = \frac{\sqrt{2}}{\pi} \partial_t \phi_{\sigma}(x, t).
\label{eq:spin_current}
\end{equation}
Equation~\eqref{eq:spin_density} shows that, in terms of physical quantities, the field $\phi_{\sigma}$ is proportional to the spin density integrated up to point $x$ at time $t$. Alternatively and equivalently, according to Eq.~\eqref{eq:spin_current}, it is proportional to the accumulated current that has flowed through point $x$ by time~$t$, added to $\phi_{\sigma}(t = 0)$. In the absence of current, $\phi_{\sigma}$ is therefore constant in time, and a nonzero current corresponds to a time-varying $\phi_{\sigma}$.
We denote the current averaged over the one-dimensional wire by
\begin{equation}
j_{\sigma}^{av}(t) = \frac{1}{L} \int_{-\frac{L}{2}}^{\frac{L}{2}} dx j_{\sigma}(x, t)
\end{equation}
and the average over both the wire length and a time period in the steady state by $\overline{j_{\sigma}}$. 

In the linear-response regime, charge transport is typically characterized by the conductance $G$, which is the change in the charge current driven by a chemical potential bias in the limit of zero bias. For spin transport, the relevant quantity is the spin conductance
\begin{equation}
G_{\sigma} = \lim_{\Delta h \to 0} \frac{\overline{j_{\sigma}}}{\Delta h}.
\end{equation}
and at finite bias, one can compute the differential conductance
\begin{equation}
\Delta G_{\sigma} = \frac{d \overline{j_{\sigma}}}{d \Delta h}.
\label{eq:differential_conductance}
\end{equation}
In the following, we report the current in units of $G_0 \Delta h_0$, where $G_0 = e^2/h$ is the conductance quantum with $e$ the elementary charge and $h$ Planck's constant. In the natural units with $e = \hbar = 1$, we have $G_0 = 1/(2\pi)$. 
For convenience, we define the quantity
\begin{equation}
\Delta h_c = \frac{2 L v_{\sigma} y_{1\perp}}{\alpha^2},
\label{eq:critical_value}
\end{equation}
which is the critical bias at the conductor-insulator transition suggested by the form of the equation of motion (see Appendices~\ref{app:derivation} and~\ref{app:critical_bias}). Here, $y_{1\perp}$ is the dimensionless coupling $y_{1\perp} = g_{1\perp}/(\pi v_{\sigma})$. We furthermore define the energy scale
\begin{equation}
\Delta h_0 = \frac{L v_F}{\alpha^2},
\label{eq:energy_scale}
\end{equation}
obtained from Eq.~\eqref{eq:critical_value} by setting $y_{1\perp} = \frac{1}{2}$ and $v_{\sigma} = v_F$. Here, $v_F$ is the Fermi velocity in the noninteracting leads.

\subsection{Solution by classical equation of motion}
\label{sec:equation_of_motion}

To obtain the spin current and spin density, we approximate the action by its stationary point along the classical path $\phi_{\sigma}(x, t)$. In other words, we formulate the classical equation of motion for $\phi_{\sigma}(x, t)$ and solve it numerically in a region of discretized space and time. This approach is the same as the one applied to a TLL wire in Ref.~\cite{Maslov_conductance1995}, albeit in that case the equation of motion can be solved analytically.
The numerical discretization of a classical field's equation of motion has previously been used to study the equilibration properties of classical integrable field theories~\cite{DeLucaMussardo2016} and to model charge transport through a periodic potential~\cite{Lebrat_band_and_correlated2018}.

The equation of motion in the bulk of the wire,
\begin{equation}
\partial_t^2 \phi_{\sigma} = v_{\sigma}^2 \partial_x^2 \phi_{\sigma} + \frac{\sqrt{2} v_{\sigma}^2 K_{\sigma} y_{1\perp}}{\alpha^2} \sin(2\sqrt{2} \phi_{\sigma}) - \frac{v_{\sigma} K_{\sigma} B}{\sqrt{2}},
\label{eq:equation_of_motion}
\end{equation}
with $B$ the magnetic-field gradient, is a second-order nonlinear differential equation obtained from the Euler-Lagrange equation, as derived in Appendix~\ref{app:derivation}.
We set the initial conditions as
\begin{equation*}
\phi(x, t = 0) = \partial_t \phi(x, t)|_{t = 0} = 0,
\end{equation*}
so that the system is at equilibrium in the initial state, there is no current and the spin density is zero. Similar to Ref.~\cite{Lebrat_band_and_correlated2018}, we implement non-reflecting boundary conditions of the form
\begin{align}
\begin{split}
\partial_t \phi(L, t) + v\partial_x \phi(L, t) &= 0, \\
\partial_t \phi(0, t) - v\partial_x \phi(0, t) &= 0.
\end{split}
\label{eq:boundary_conditions}
\end{align}
These boundary conditions correspond to the continuity equation for a conserved quantity, which in this case is the field $\phi_{\sigma}$. The current is transported into and out of the wire the same way as it would for infinitely long leads. In particular, there is no reflection of waves at the boundaries. 
Transparent boundary conditions for the sine-Gordon model were studied in detail in Ref.~\cite{SabirovMatrasulov2022}. 

\begin{figure}
\centering
\includegraphics[width=0.82\linewidth]{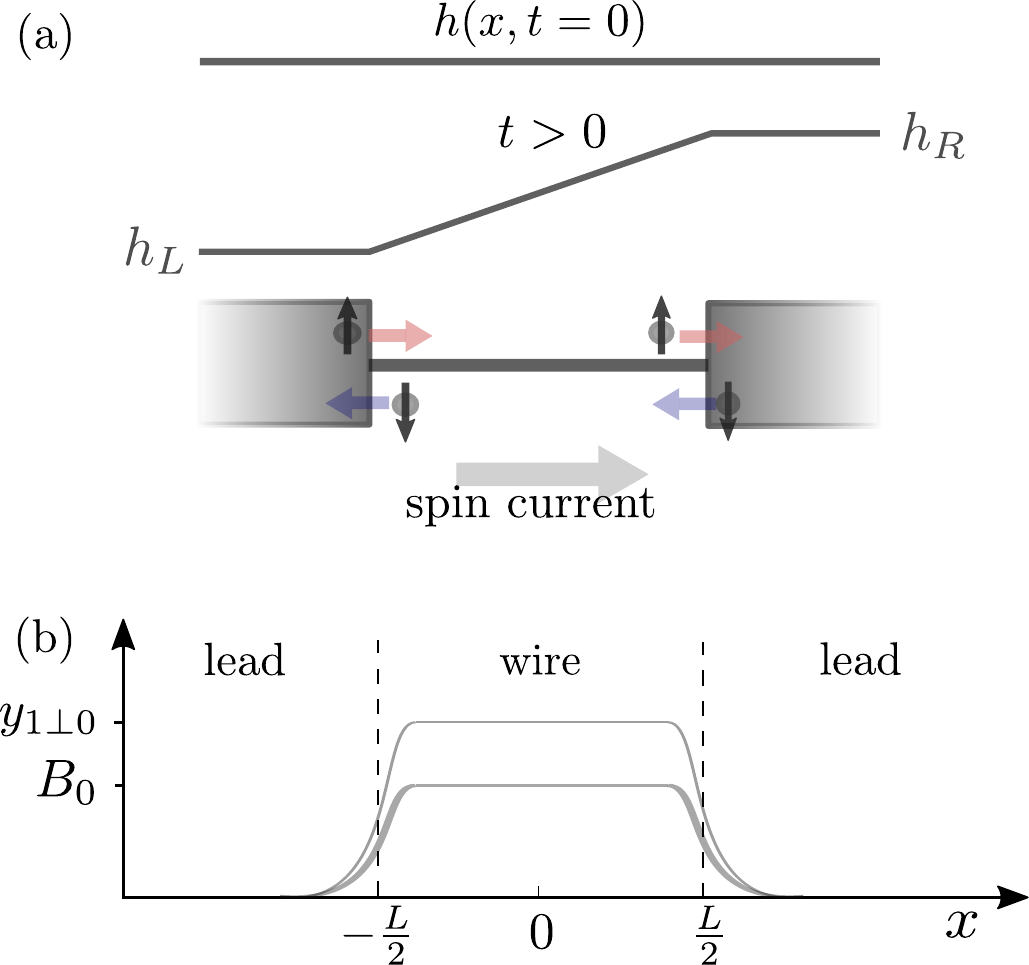}
\caption{(a) The magnetic field is quenched from zero to nonzero at $t = 0$. For $t > 0$, it is given by Eq.~\eqref{eq:magnetic_field}. As a result, the spins rearrange and a spin current is generated. We choose the sign of the magnetic field so that a positive spin current $j_{\sigma} = j_{\uparrow} - j_{\downarrow}$ flows from left to right.
(b) The parameters $y_{1\perp}(x)$ and $B(x)$ of Eq.~\eqref{eq:equation_of_motion} are set to zero within the leads and change smoothly into the values $y_{1\perp 0}$, $B_0$ within the wire.
}
\label{fig:field}
\end{figure}

A spin current is driven by a spin bias, i.e. a difference in the magnetic fields in the left and right reservoir, $\Delta h = h_L - h_R$. 
To obtain a positive spin current flowing from left to right, we impose a negative magnetic field in the left lead and a positive one in the right lead, switched on at time $t = 0$. This leads to a flow of spin-up fermions towards the right lead and spin-down towards the left, as illustrated in Fig.~\ref{fig:field}.
We denote the magnetic-field gradient by $B(x)$, so that $h(x)$ depends on position as
\begin{equation}
h(x) = h_L - \int_{-\frac{L}{2}}^x B(x^\prime) d x^{\prime},
\end{equation}
where $h_L = h(-L/2)$. We impose a constant field gradient within the wire and 
assume, analogous to Ref.~\cite{Maslov_conductance1995}, that $h$ has a constant value in each lead and changes linearly within the wire, 
\begin{equation}
h(x) = h_L - B_0 \left(x + \frac{L}{2} \right).
\label{eq:magnetic_field}
\end{equation}
Here, $B_0 = \Delta h/L$.

The equation of motion is solved numerically by discretizing the space and time coordinates, as detailed in Appendix~\ref{app:discretization}.
Both $y_{1\perp}(x)$ and $B(x)$ have a constant value within the wire and are zero in the leads. 
For numerical stability, these parameters are changed smoothly at the boundaries as 
\begin{equation}
f(x) = \frac{f_0}{2} \left[ \mathrm{erf}\left( \frac{x + \frac{L}{2}}{w} \right) + \mathrm{erf}\left( \frac{-x + \frac{L}{2}}{w} \right) \right]
\label{eq:sigmoid}
\end{equation}
where $f = y_{1\perp}, B$. This function is drawn schematically in Fig.~\ref{fig:field}(b). The parameter $w$ controls the width of the boundary regions. We choose $w$ larger than the spatial discretization but much smaller than the wire length and check that the choice of $w$ does not change the results qualitatively. 
In the leads, Eq.~\eqref{eq:equation_of_motion} has the form of a wave equation $\partial_t^2 \phi_{\sigma} = v_F^2 \partial_x^2 \phi_{\sigma}$. In the following, we use $y_{1\perp}$ to refer to the constant value $y_{1\perp 0}$ within the wire.

\section{General solution}
\label{sec:general_solution}

In this section, we characterize the solution for generic parameters $K_{\sigma}$, $v_{\sigma}$ and $y_{1\perp}$ not specific for any microscopic model. We discuss the general features of the solution when the spin bias is varied. 
We set here the velocity and the Luttinger parameter constant for simplicity, i.e. equal in the wire and leads, $K_{\sigma} = K_L = 1$ and $v_{\sigma} = v_F$. In Sec.~\ref{sec:different_K_v}, consider different $K_{\sigma}$ and $v_{\sigma}$ in the wire and in the leads, and in Sec.~\ref{sec:experimental}, the parameters are set by the experimental parameters of Ref.~\cite{Krinner_spin_and_particle2016}.

\subsection{Initial time evolution}
\label{sec:initial}

\begin{figure}
\centering
\includegraphics[width=\linewidth]{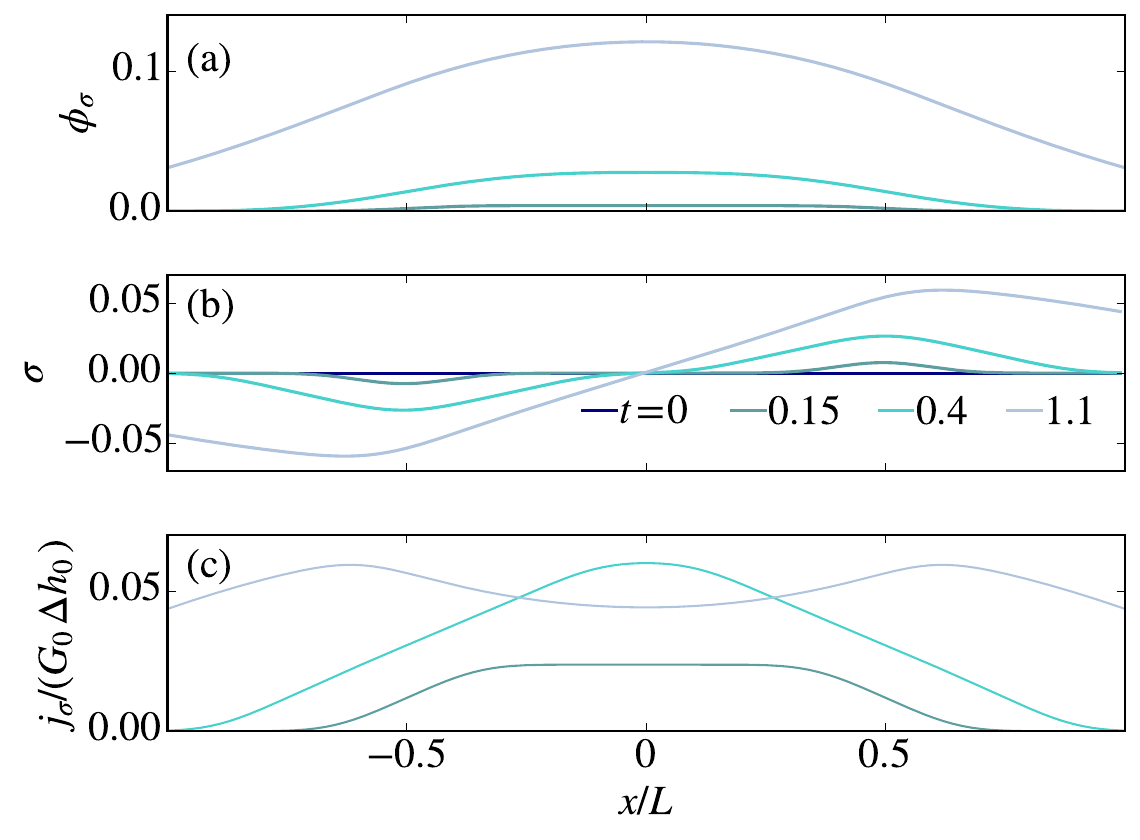}
\caption{The spatial profiles of (a) the field $\phi_{\sigma}$, (b) the spin density, and (c) the spin current, at fixed times during the initial time evolution. The wire region is within $x \in [-L/2, L/2]$ while $|x| > L/2$ corresponds to the leads. Here, we set $\Delta h/\Delta h_c = 0.5$, $v_{\sigma} = v_F$, and time is in units of $L/v_F$.}
\label{fig:profiles}
\end{figure}

We analyze here the numerical solution $\phi_{\sigma}(x, t)$, the spin density $\sigma(x, t)$, and the spin current $j_{\sigma}$. 
At time $t=0$, the magnetic field is quenched from zero to the profile of Eq.~\eqref{eq:magnetic_field}. This induces a spin current, illustrated in Fig.~\ref{fig:field}(a), as fermions with spin up start to move towards the right lead and spin-down fermions towards the left lead. A deformation of the spin density originates at the contacts where the reorganization of spins is possible:
Figure~\ref{fig:profiles}(b) shows how the spin density is depleted around the left contact and accumulates around the right one.

The spin current profile is plotted at different times in Fig.~\ref{fig:profiles}(c). 
Driven by the spin bias, the spin current grows to its maximum value, and both the current and the spin density deformations spread into the reservoirs as more particles are involved in transport. 
The linearly changing magnetic field inside the wire leads to a spin density that changes linearly in this region. 
The transport of spin within the wire is impeded due to the interactions giving rise to a spin gap. As the spin density is not sufficiently replenished by the reservoirs, the current in the wire starts to decrease. This evolution is different from a Luttinger liquid wire, where transport is ballistic and the current evolves into a uniform steady-state distribution (see Appendix~\ref{app:LL_wire}).
The initial time evolution of the spin density and spin current is qualitatively similar in both the conducting and insulating parameter regimes. In the insulating one, the current subsequently decays to zero, whereas in the conducting regime, the competition of the spin gap and the spin bias gives rise to a periodic reduction and increase, as discussed below.

\subsection{Transition from insulator to conductor}
\label{sec:transition}

\begin{figure}
\centering
\includegraphics[width=\linewidth]{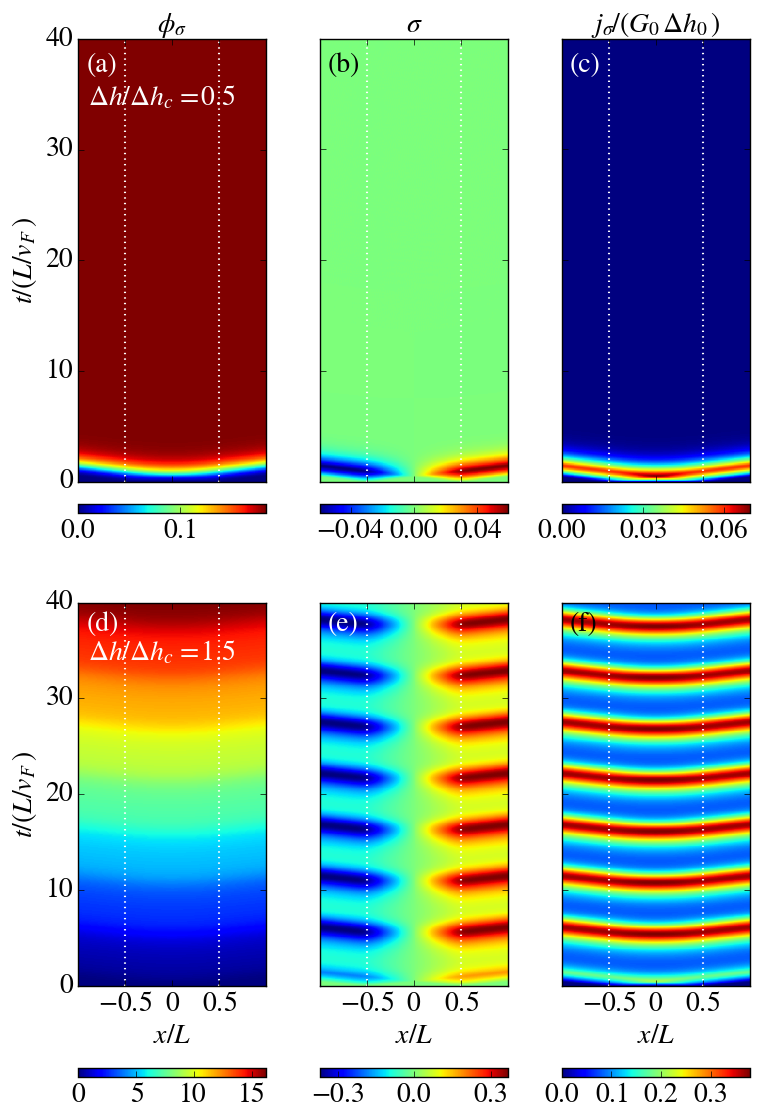}
\caption{The field $\phi_{\sigma}$ (a, d), the spin density $\sigma$ (b, e), and the spin current $j_{\sigma}$ (c, f) as functions of position $x$ and time $t$. The upper row with $\Delta h/\Delta h_c = 0.5$ corresponds to an insulating steady state and the lower row with $\Delta h/\Delta h_c = 1.5$ corresponds to a conducting one. We set $v_{\sigma} = v_F$. The wire region is between the white dotted vertical lines.}
\label{fig:phi}
\end{figure}

After the initial transient time evolution, the system evolves into a steady state. In Fig.~\ref{fig:phi}, we plot the field $\phi_{\sigma}$, spin density, and spin current as functions of position and time.
The initial time evolution shows a ``light cone'' structure where the spin density deformation and current originating at the contacts propagate into the leads. As the boundary conditions are non-reflecting, there are no boundary effects, and the system corresponds to a wire coupled to infinite leads similar to Ref.~\cite{Maslov_conductance1995}.
After the initial transient time evolution, the system reaches a steady state where the current averaged over a time period is constant. Two kinds of steady states occur: an insulating and a conducting one. 
In Fig.~\ref{fig:phi}, the transition takes place when $\Delta h$ exceeds its critical value given by Eq.~\eqref{eq:critical_value}. 

At equilibrium and in the thermodynamic limit, a phase transition occurs between a spin-gapped and a TLL phase when the magnetic field exceeds the spin gap~\cite{JaparidzeNersesyan1979, Giamarchi_one_dimension2003}. 
In a wire coupled to leads, however, the finite length of the wire plays a role and one may rather expect a Landau-Zener-type transition~\cite{Landau1932, Zener1932, Zener1934}. In the charge transport through a Mott insulator wire, the Landau-Zener dielectric breakdown was found to occur at a threshold electric field $E_{\text{th}} \sim \Delta^2/W$, where $\Delta$ is the charge gap and $W$ the bandwidth~\cite{OkaAoki2003, OkaAoki2005, OkaAoki2010, HeidrichMeisnerDagotto2010, TanakaYonemitsu2013, TakasanKawakami2019}. For spin transport, we may expect a threshold magnetic-field gradient $B_{\text{th}} \sim \Delta_{\sigma}^2/v_{\sigma}$, so that the critical spin bias $\Delta h_{\text{LZ}} = L \Delta_{\sigma}^2/v_{\sigma}$ gives an estimate for the spin gap in the system.
We note that the critical bias in the classical equation of motion may nevertheless not give an accurate estimate of the spin gap of a given microscopic quantum system, since quantum fluctuations, not taken into account here, would lead to a renormalization of the parameters. We come back to this point in Secs.~\ref{sec:experimental} and~\ref{sec:discussion}.

\begin{figure}
\centering
\includegraphics[width=\linewidth]{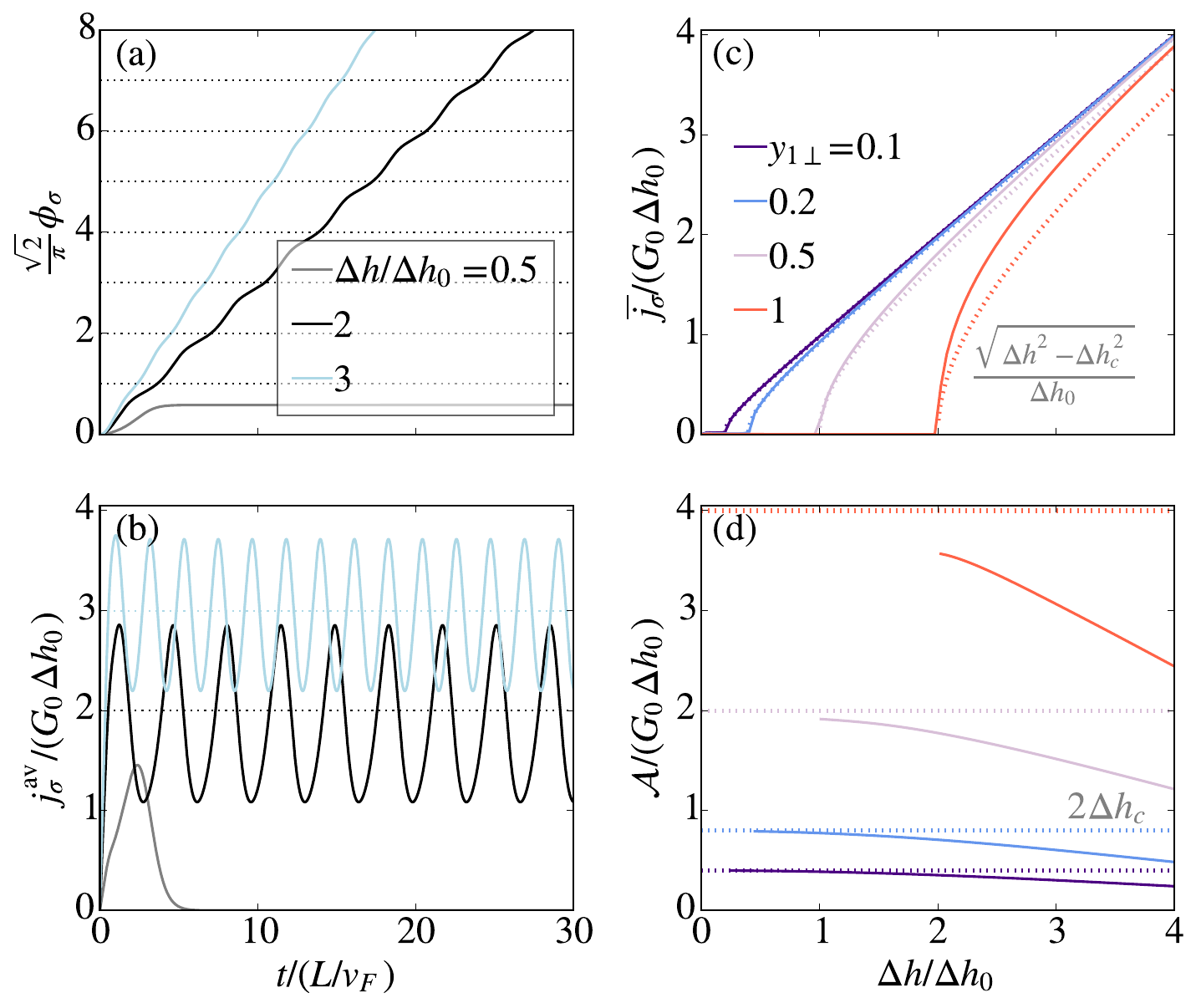}
\caption{(a) The solution $\phi_{\sigma}$ as a function of time, averaged over the wire length. For a spin bias below the critical value ($\Delta h/\Delta h_c = 0.5$), the steady-state solution is constant, while above the critical bias, $\phi_{\sigma}$ grows linearly with a modulation in steps of $\pi/\sqrt{2}$. Here, $v_{\sigma} = v_F$ and $y_{1\perp} = 0.5$ so that $\Delta h_c = \Delta h_0$. 
(b) The spin current $j_{\sigma}^{\text{av}}$ corresponding to panel~(a) decays to zero when $\Delta h < \Delta h_c$ and oscillates with a nonzero time average when $\Delta h > \Delta h_c$. The dotted lines show the limiting value $G_0 \Delta h$ for $\Delta h \gg \Delta h_c$. 
(c) The steady-state time-averaged spin current $\overline{j_{\sigma}}$ as a function of the bias. The solid lines are the solution of the equation of motion for different couplings $y_{1\perp}$ and the dotted lines show Eq.~\eqref{eq:critical_bias}. 
(d) The steady-state oscillation amplitude $\mathcal{A}$ as a function of the bias. Close to the critical bias, the amplitude approaches $2 G_0 \Delta h_c$.}
\label{fig:solitons}
\end{figure}

For $\Delta h < \Delta h_c$, as in Figs.~\ref{fig:phi}(a)--(c), the transient oscillations are damped and $\phi_{\sigma}$ reaches a constant value in the steady state where the spin density and spin current are zero. 
This corresponds to an insulating state of the wire. For $\Delta h > \Delta h_c$, the steady state is characterized by persistent oscillations with constant frequency and amplitude. The field $\phi_{\sigma}$ shows a linear increase modulated by oscillations, and $j_{\sigma}$ oscillates with a nonzero average value that is constant in time. The wire is therefore in a conducting steady state. 
This transition is seen in Fig.~\ref{fig:solitons}(c), which shows $j_{\sigma}$ averaged over several oscillation periods in the steady state, as a function of $\Delta h$. We have fixed the parameters so that the transition occurs at $\Delta h = \Delta h_0$ when $v_{\sigma} = v_F$, in particular, we set here $y_{1\perp} = \frac{1}{2}$. Conversely, fixing $\Delta h$ and varying $y_{1\perp}$ produces the same phase transition.
Physically, the existence of a steady state with a constant time average of $j_{\sigma}$ is enabled by the infinite reservoirs to which the wire is coupled, providing an endless supply of particles and unchanged by the spin-polarized current that flows between them.

In Figs.~\ref{fig:profiles}--\ref{fig:varying_Ksigma}, the wire length is set to $L = 1$ and the regions designated as the leads have length $L/2$, so that the solution is computed for $x \in [-1, 1]$. Extending the leads beyond $|x| = 1$ does not change the solution, up to numerical errors on the order of $10^{-4}$ for the discretization used here (see Appendix~\ref{app:discretization}).

\subsection{Oscillating solution}
\label{sec:oscillation}

In the conducting state, the oscillating solution may be thought of in terms of instanton solutions of the classical field~\cite{Rajaraman1982, Giamarchi_one_dimension2003}: The spin current is mediated by a succession of instanton excitations. 
An instanton transfers the field $\phi_{\sigma}$ from one minimum of the cosine to the next, which for Eq.~\eqref{eq:equation_of_motion} corresponds to a change by $\pi/\sqrt{2}$, seen in Fig.~\ref{fig:solitons}(a). 
This has similarity to the equilibrium situation where a sufficiently high magnetic field induces a phase transition from a spin-gapped phase into a Luttinger liquid with a finite density of solitons~\cite{Haldane1982} 
dispersed with the distance $\sim v_{\sigma}/h$ between them~\cite{Giamarchi_one_dimension2003}. We find that $\phi_{\sigma}$ grows in steps of $\pi/\sqrt{2}$ with a time period $\tau \sim \Delta h^{-1}$. 
In the case of charge transport in a charge-density-wave wire, described by the sine-Gordon model, the conductivity~\cite{Maki1987} and DC conductance~\cite{MoriFukuyama1997, Ponomarenko_Mott_insulator1998, Ponomarenko_spin_gap_insulators2000, KriveOxman2000} were discussed previously in terms of soliton and instanton transport.

The spin current corresponding to Fig.~\ref{fig:solitons}(a) is shown in Fig.~\ref{fig:solitons}(b). For $\Delta h/\Delta h_c = 0.5$, the current has an initial transient peak and then relaxes to zero, while for $\Delta h > \Delta h_c$, $j_{\sigma}$ oscillates with a nonzero average and a fixed frequency and amplitude in the steady state. The average slope of $\pi/\sqrt{2} \phi_{\sigma}$ is equal to the time-averaged value $\overline{j_{\sigma}}$. For large bias $\Delta h \gg \Delta h_c$, the time average approaches $G_0 \Delta h$, marked with dotted lines in Fig.~\ref{fig:solitons}(c).
Note that the average slope of $\pi/\sqrt{2} \phi_{\sigma}$ is also the inverse oscillation period, and therefore the oscillation frequency $\omega = 2\pi/\tau$ is connected to the time-averaged current as $\omega = 2\pi \overline{j_{\sigma}}$.

We plot $\overline{j_{\sigma}}$ in Fig.~\ref{fig:solitons}(c) as a function of $\Delta h$ for different coupling constants $y_{1\perp}$.
To ensure that the system has reached a steady state, we extract the time-averaged quantities from the later half of a time evolution up to $t/(L/v_F) = 250$.
For the values $y_{1\perp} \leq 0.5$, the numerically evaluated average spin current is closely reproduced by the expression 
\begin{equation}
G_0 \sqrt{\Delta h^2 - \Delta h_c^2},
\label{eq:critical_bias}
\end{equation}
while for the stronger coupling $y_{1\perp} = 1$ there is a larger deviation. This is similar to the expression for the voltage $V = R\sqrt{I_{\text{ext}}^2 - I_c^2}$ induced by a DC current drive $I_{\text{ext}}$ in an overdamped Josephson junction with critical current $I_c$ and resistance $R$~\cite{Tinkham1996}. This correspondence is not obvious since the overdamped Josephson junction is described by a first-order differential equation and does not contain a spatial derivative as in Eq.~\eqref{eq:equation_of_motion}. 

The oscillation amplitude $\mathcal{A}$ shown in Fig.~\ref{fig:solitons}(d) is larger for larger coupling $y_{1\perp}$ and decays with $\Delta h$, although we do not find a simple functional dependence.  
In overdamped Josephson junctions, 
the induced voltage is a periodic series of pulses with a maximum value $2 I_c R$ for $I_{\text{ext}} \gtrsim I_c$~\cite{Tinkham1996}. Comparing to Eq.~\eqref{eq:critical_bias} suggests an amplitude $2G_0 \Delta h_c$ for $\overline{j_{\sigma}}$. We find that close to the critical bias, the amplitude approaches this value, marked by the dotted lines in Fig.~\ref{fig:solitons}(d).

\subsection{Different $K_{\sigma}$ in the wire and in the leads}
\label{sec:different_K_v}

In the previous sections, we considered $K_{\sigma}$ 
equal in the wire and in the leads. Here, we investigate the effects of a different $K_{\sigma}$ 
in the wire than in the leads when the wire is described by the sine-Gordon model and the leads are Luttinger liquids with $y_{1\perp} = 0$. We present both the case of noninteracting leads and interacting TLL leads.
The parameter $K_{\sigma}$ changes smoothly at the boundary of the wire and leads, similar to $y_{1\perp}(x)$ and the magnetic-field gradient $B(x)$ shown in Fig.~\ref{fig:field}.
For an interacting TLL wire, it is known that the conductance is determined by the Luttinger parameter of the leads, $G = K_L G_0$, instead of that of the wire~\cite{Safi_transport1995, Maslov_conductance1995, Ponomarenko_conductance1995}. 
This was shown to be true also for a spatially varying velocity and Luttinger parameter within the wire, given that they approach the constant parameters in the leads smoothly~\cite{ThomaleSeidel2011}.
We therefore investigate whether the same result is found for the differential conductance of the sine-Gordon wire at $\Delta h \gg \Delta h_c$ where the current-bias curve is linear. 

\begin{figure}
\centering
\includegraphics[width=\linewidth]{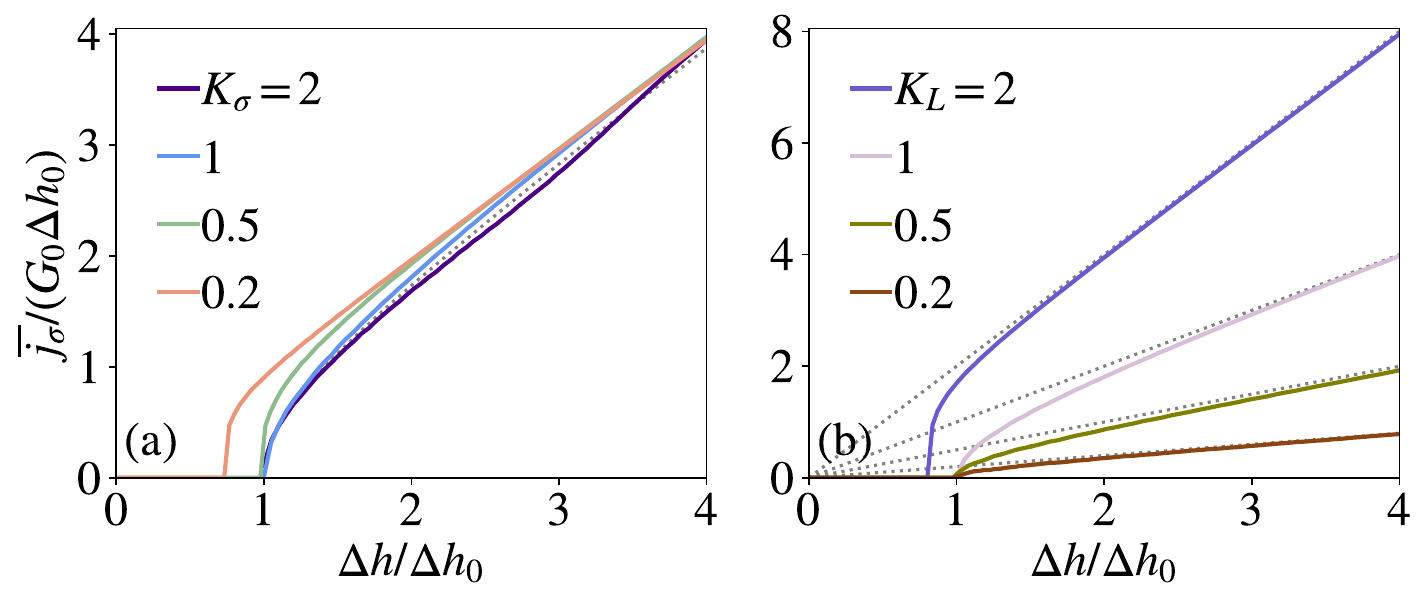}
\caption{
(a) The time-averaged spin current as a function of the spin bias for varying $K_{\sigma}$ within the wire. The other parameters are fixed as $y_{1\perp} = 0.5$, $K_L = 1$, $v_{\sigma} = v_F$. For $K_{\sigma} \gtrsim 0.5$, the current is closely reproduced by Eq.~\eqref{eq:critical_bias} marked by the dotted line, and approaches $G_0 \Delta h$ at large bias for all $K_{\sigma}$.
(b) The time-averaged spin current as a function of the spin bias for varying $K_L$ with $K_{\sigma} = 1$ within the wire. The other parameters are as in panel~(a). The spin current approaches $K_L G_0 \Delta h$ at large bias, indicated by the dotted lines.
}
\label{fig:varying_Ksigma}
\end{figure}

Figure~\ref{fig:varying_Ksigma}(a) shows the time-averaged steady-state spin current as a function of bias for different values of $K_{\sigma}$ in the wire, while the leads are noninteracting and the spin velocity is fixed to $v_{\sigma} = v_F$ everywhere. The critical bias is $\Delta h_c = \Delta h_0$ given by Eq.~\eqref{eq:critical_value} for all but the smallest value of $K_{\sigma}$ shown here. For $K_{\sigma} = 0.2$, it is reduced in a nontrivial way. 
This may be related to whether the correlation length $\xi_{\sigma} \sim v_{\sigma}/\Delta_{\sigma}$ is larger or smaller than the wire length $L$; in Mott insulator wires, the threshold voltage for the Landau-Zener breakdown was found to have a different dependence on the charge gap in these two cases~\cite{TanakaYonemitsu2013} (see also Appendix~\ref{app:critical_bias}). 
The current at large bias $\Delta h \gg \Delta h_c$ approaches the line $G_0 \Delta h$ signifying ballistic transport, similar to the Luttinger liquid wire. In Fig.~\ref{fig:varying_Ksigma}(b), we fix $K_{\sigma} = 1$ within the wire while the leads are interacting Luttinger liquids with $K_L \neq 1$. In this case, we observe that the critical bias is reduced for the largest value $K_L = 2$. The differential conductance $K_L G_0$ is recovered in the limit $\Delta h \gg \Delta h_c$.

\section{Experimental parameters}
\label{sec:experimental}

While the magnetic field configuration of Fig.~\ref{fig:field} could be realized by applying an external magnetic field, we discuss, in this section, a different experimental situation where the spin current originates from an initial spin population imbalance between the two reservoirs. 
The imbalance can be modeled by including a spin bias $\Delta h$, analogous to the case of charge transport due to a chemical potential bias in Ref.~\cite{Lebrat_band_and_correlated2018}.
In particular, we discuss the parameter regime relevant for the cold-atom experiment of Ref.~\cite{Krinner_spin_and_particle2016}, where spin and particle conductances  were measured in both conducting and insulating parameter regimes.
We compare the predicted oscillation period of the spin current to the reported time scales in order to estimate whether the oscillation dynamics could be observable in that setup. 
In the experiment, two atom cloud reservoirs, initially separated, are prepared with a spin population imbalance. The imbalance translates to a spin bias via the equation of state.
Connecting the reservoirs allows them to exchange particles, generating a spin current. The spin imbalance decays as a function of time, which is different from the constant magnetic-field gradient considered here. We nevertheless estimate the expected oscillation period and compare it to the typical experimental transport time scales to determine whether these oscillations could be observed in the short-time dynamics. While a constant bias occurs naturally in solid-state devices, initializing and measuring spin currents in the solid state is challenging. The cold-atom setup therefore offers an interesting route to studying spin transport.

In Refs.~\cite{Krinner_spin_and_particle2016, Lebrat_band_and_correlated2018}, the reservoirs are attractively interacting but are at a finite temperature. If the temperature is larger than the spin gap, the effects of pairing should be negligible and we can assume that noninteracting leads are a reasonable description. We therefore consider interactions only in the wire.
The interactions between the fermionic atoms in different spin states arise from s-wave scattering, which microscopically can be modeled by spin-rotation-invariant contact interactions. Due to spin-rotation invariance, the parameters $K_{\sigma}$ and $y_{1\perp}$ are related through $y_{1\perp} = 2 \left( K_{\sigma}^2 - 1 \right)/\left( K_{\sigma}^2 + 1 \right)$. The coupling $y_{1\perp}$ is determined by the scattering length $a$ and the particle density $\rho_0$ in the wire as described in Appendix~\ref{app:experimental}. 

\begin{figure}
\centering
\includegraphics[width=\linewidth]{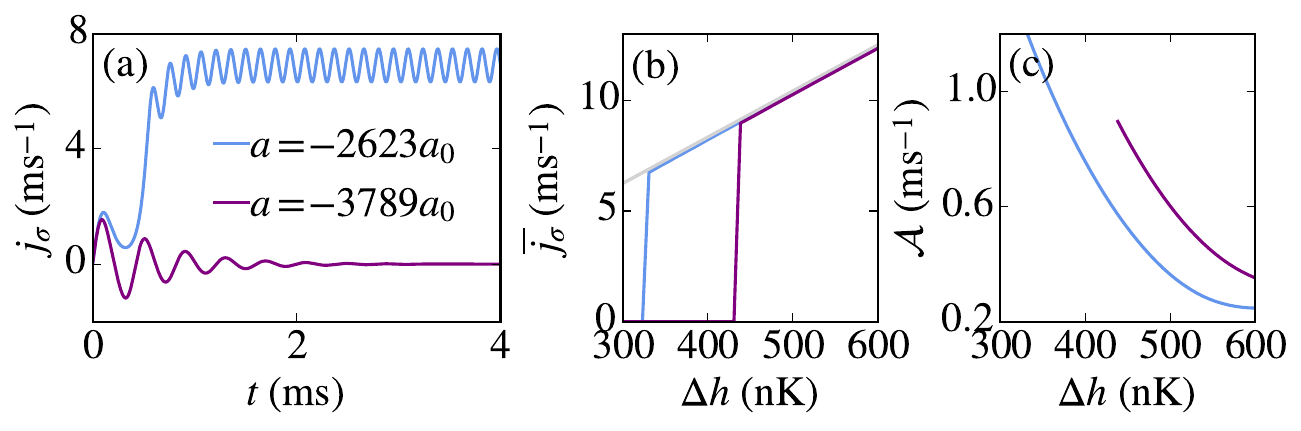}
\caption{(a) The spin current $j_{\sigma}$ averaged over the wire length, multiplied by $2\pi$, as a function of time for two different scattering lengths. The spin bias is $340$~nK, for which the spin current oscillates when $a = -2623 a_0$ and relaxes to zero when $a = -3789 a_0$. The particle density in the wire is set to $\rho_0 = 0.8/\mu\text{m}$ and the wire length is $L = 6$ $\mu$m.
(b) The time-averaged spin current in the steady state is close to $G_0 \Delta h$ at large bias. 
(c) The amplitude $\mathcal{A}$ of the oscillation decays with $\Delta h$.}
\label{fig:experimental}
\end{figure}

We compute the time evolution of the spin current using the smallest and largest interaction strengths quoted in Refs.~\cite{Krinner_spin_and_particle2016, VisuriGiamarchi2020}.
For the bias $\Delta h = 340$ nK used in Fig.~\ref{fig:experimental}(a), the weaker interaction with $a = -2623 a_0$ leads to an oscillating steady state with a nonzero average spin current. Here, $a_0$ is the Bohr radius $a_0 = 5.29 \times 10^{-11}$ m. For the stronger interaction ($a = -3789 a_0$), the current in the steady state is zero, and an oscillating steady state is reached with a larger bias.
Figure~\ref{fig:experimental}(b) shows the time-averaged spin current in the steady state as a function of the bias.
For the parameters used here, the current changes seemingly discontinuously from zero to a finite value when the spin bias exceeds the critical value. We expect that this discontinuity would be smoothed out if the effects of a finite temperature were taken into account. 
The average spin current approaches the value $G_0 \Delta h$ expected for ballistic transport, consistent with the results of the previous section.

The current shown in Fig.~\ref{fig:experimental}(b) is of similar order of magnitude as the particle currents reported in Ref.~\cite{Krinner_spin_and_particle2016}, where the atom number imbalance changes by around $10^4$ within $1$ s. 
The oscillation amplitude is of similar order of magnitude as the reported changes in atom numbers between data points.
The initial spin bias is on average $0.24\mu$~\cite{Krinner_spin_and_particle2016}, while the chemical potential $\mu$ is quoted as $360$ nK in Ref.~\cite{HusmannBrantut2015}. A spin bias of order $0.24\mu \approx 90$ nK is smaller than the bias $\sim 340$ nK at which the wire becomes conducting for the smaller interaction in Fig.~\ref{fig:experimental}(b). 
On the other hand, the classical equation of motion may not reflect the true value of the spin gap, as its validity here is limited to the strongly attractive regime. 
A more precise value can be found by solving the spin gap directly for the Gaudin-Yang model using Bethe ansatz equations~\cite{Fuchs_exactly_solvable2004}. 
A rough order-of-magnitude estimate for the critical bias can then be obtained as $\Delta h_{\text{LZ}} = L \Delta_{\sigma}^2/(\hbar k_B v_{\sigma})$, as discussed in Sec.~\ref{sec:transition}. For the particle density $\rho_0 = 0.8/\mu$m and the scattering lengths $a = -2623 a_0$ and $-3789 a_0$, we get $\Delta h_{\text{LZ}} \approx 32$ and $180$ nK, respectively (see Appendix~\ref{app:experimental}).
The oscillating current could then be present at spin biases similar to the ones used in Ref.~\cite{Krinner_spin_and_particle2016}. The oscillation period would be around $0.5$ ms for $\Delta h = 90$ nK. For biases below the experimental temperature $60$ nK~\cite{Krinner_spin_and_particle2016}, the oscillations would likely be blurred out by thermal fluctuations.
We also note here another point of experimental relevance: when the short-distance cutoff $\alpha$ is set as the inter-particle distance $\rho_0^{-1}$, the critical spin bias obtained from Eq.~\eqref{eq:critical_bias} is proportional to $\rho_0^2$. In contrast, the Gaudin-Yang spin gap decays with the particle density, as does the critical bias $\Delta h_{\text{LZ}}$ estimated from this gap ($v_{\sigma}$ increases with $\rho_0$). The density dependence of the critical bias could therefore help to identify the appropriate theoretical description of the system.

The exponential decay of the spin imbalance has a time constant that is related to the transport time used in the measurements in Ref.~\cite{Krinner_spin_and_particle2016}. We use here the transport time $t_{\text{tr}} = 4$ s as a reference time scale. The oscillation period $0.5$ ms is orders of magnitude smaller than the characteristic transport time, so that several oscillation periods could be realized before the spin bias decays significantly. 
As the bias decays in time, the oscillation period, inversely proportional to the bias, would, in a simple picture, increase as a function of time. 
The decay curves can be used to reconstruct the full current-bias characteristics~\cite{HusmannBrantut2015}, and the presence of a critical bias should appear as a nonzero long-time asymptote of the spin imbalance.

\section{Discussion}
\label{sec:discussion}

Solving the classical equation of motion for $\phi_{\sigma}$ amounts to approximating quantum field theory path integrals by the classical path along which the action is stationary. This approximation can be expected to be most accurate when the quantum fluctuations of $\phi_{\sigma}$ have a negligible contribution.  
For the sine-Gordon model, this corresponds to the regime where the cosine term in the Hamiltonian is relevant in renormalization, i.e. the renormalization flows to strong coupling (see for example Ref.~\cite{VisuriGiamarchi2020}), fermions with opposite spin are paired and the spin gap is nonzero. In that case, $\phi_{\sigma}$ is fixed to the value that minimizes the cosine, and fluctuations around this minimum are negligible. For the cold-atom experiment with spin-rotation invariant contact interactions discussed in Sec.~\ref{sec:experimental}, this is the strongly attractive regime. In the weakly attractive case, one can expect deviations from the classical result, as well as in the repulsive one, where the cosine term is irrelevant due to strong quantum fluctuations. 

Quantum effects are known to lead to a mass renormalization of sine-Gordon solitons and antisolitons~\cite{Rajaraman1982}. It would be interesting to study whether this has consequences on the critical bias for the spin transport mediated by instanton excitations, as 
we find the spin gap to be overestimated by the solution of the classical equation of motion compared to the exact spin gap in the Gaudin-Yang model.
Taking into account quantum fluctuations might explain this deviation if the possibility of quantum tunneling of $\phi_{\sigma}$ from one minimum of the cosine to the next reduces the critical bias. 

Furthermore, considering the effects of a finite temperature would be interesting in connection with experiments. In Ref.~\cite{Lebrat_band_and_correlated2018}, to make comparisons with experimental data, the effects of finite-temperature reservoirs were modeled by stochastic noise at the boundaries. In statistical physics, one commonly includes both a friction term proportional to $\partial_t \phi_{\sigma}$ and a stochastic fluctuation, which at thermal equilibrium are connected via the fluctuation-dissipation theorem and determine the temperature. We do not include these terms in the equation of motion, and our model therefore can be thought to describe an isolated quantum system at a finite energy density. In Ref.~\cite{Lebrat_band_and_correlated2018}, a finite temperature is found to contribute to spin transport as thermal excitation energies exceed the spin gap. Here, we expect that accounting for a finite temperature by including a stochastic noise would lead to a smoother change of the current around the critical bias and smearing out of the current oscillations.

\section{Conclusions}
\label{sec:conclusions}

In an interacting wire coupled to noninteracting or Luttinger liquid leads, the spin current displays a transition from a spin-insulating to a spin-conducting phase when the spin bias exceeds its critical value. 
In the conducting phase, the spin current oscillates with a fixed frequency given by the spin bias, and the solution has similar features to a damped Josephson junction driven by a DC current. When the Luttinger parameter is different in the wire than in the leads, we recover a differential conductance proportional to the Luttinger parameter of the leads in the limit of large spin bias.
We furthermore estimate the oscillation period in physical units and compare it to characteristic time scales reported for the cold atom transport experiment of Refs.~\cite{Krinner_spin_and_particle2016, Lebrat_band_and_correlated2018}. The oscillation period is orders of magnitude smaller than the typical transport time, suggesting that the oscillation could be observable in the short-time dynamics.

There is a multitude of systems where the low-energy physics is described by the sine-Gordon model.
Observing dynamical effects resulting from the underlying sine-Gordon model in spin transport would be particularly intriguing as the cosine term arises purely from interactions. 
A similar model can be realized for charge transport in a periodic potential~\cite{Lebrat_band_and_correlated2018}, although in this case there are coupling terms between the spin and charge sectors. The spin sector, in the absence of external potentials, is therefore a cleaner realization of the sine-Gordon model and could offer an interesting (quantum simulation) platform to study its dynamics. The solution of the classical equation of motion considered here could further be used to examine the effects of inhomogeneities in the sine-Gordon parameters. Beyond the classical solution, it would be interesting to investigate the effects of quantum fluctuations on the transport dynamics.

\begin{acknowledgments}
The author is grateful to M. Filippone, T. Giamarchi, P. Gri\v{s}ins, and C. Kollath for helpful discussions; M. Lebrat, L. Foini, and S. Uchino for comments on the paper; and the Lithium team at ETH Zurich (T. Esslinger, P. Fabritius, M.-Z. Huang, J. Mohan, M. Talebi, and S. Wili) for inspiring collaborations on related topics. 
The author acknowledges funding from the Deutsche Forschungsgemeinschaft, in particular under Project No. 277625399---TRR 185 (B3), Project No. 277146847---CRC 1238 (C05) and Germany’s Excellence Strategy---Cluster of Excellence Matter and Light for Quantum Computing (ML4Q) EXC2004/1---Project No. 390534769. 
\end{acknowledgments}

\appendix

\section{Derivation of the equation of motion}
\label{app:derivation}

The largest contribution to path integrals is given by the classical path $\phi_{\text{cl}}$ along which the action is stationary,
\begin{equation}
\frac{\delta S}{\delta \phi}\Big|_{\phi = \phi_{\text{cl}}} = 0.
\label{eq:stationary}
\end{equation}
The real-time action $S = \int dx \int dt \mathcal{L}(\phi, \partial_t \phi, \partial_x \phi)$ is written in terms of the Lagrangian density $\mathcal{L}$, connected to the Hamiltonian density as
\begin{equation}
\mathcal{L} = \Pi \partial_t \phi - H.
\end{equation}
We rewrite the canonical momentum $\Pi = \frac{\hbar}{\pi}\partial_x \theta$ using the duality relation 
\begin{equation}
\Pi = \frac{\hbar}{\pi K v}\partial_t \phi, 
\end{equation}
where $\hbar$ is included for completeness. 
The Lagrangian for the forced sine-Gordon model, corresponding to Eq.~\eqref{eq:total_hamiltonian}, is then
\begin{align}
\begin{split}
\mathcal{L} = \frac{\hbar}{2 \pi K} &\left(\frac{1}{v_{\sigma}} \phi_t^2 - v_{\sigma}\phi_x^2 \right) \\
&-\frac{2\pi \hbar v_{\sigma} y_{1\perp}}{(2\pi\alpha)^2} \cos(2\sqrt{2} \phi) - \frac{h(x)}{\sqrt{2}\pi} \phi_x.
\end{split}
\end{align}
Here, we shorten the notation to $\partial_x \phi = \phi_x$, $\partial_x^2 \phi = \phi_{xx}$.

From Eq.~\eqref{eq:stationary}, one obtains the classical equation of motion, or the Euler-Lagrange equation, as
\begin{equation*}
\frac{\delta \mathcal{L}}{\delta \phi} = \frac{\partial}{\partial x} \left( \frac{\delta \mathcal{L}}{\delta \phi_x} \right) + \frac{\partial}{\partial t} \left( \frac{\delta \mathcal{L}}{\delta \phi_t} \right).
\end{equation*}
Taking into account the spatial variation of $v_{\sigma}$, $K_{\sigma}$ and $y_{1\perp}$, we find the different terms as
\begin{align*}
\frac{\partial}{\partial x} \frac{\delta \mathcal{L}}{\delta \phi_x} 
&= -\frac{\partial}{\partial x} \left[\frac{\hbar v_{\sigma}(x)}{\pi K_{\sigma}(x)} \phi_x 
+ \frac{h(x)}{\sqrt{2} \pi} \right] \\
&= -\frac{\hbar v_{\sigma}(x)}{\pi K_{\sigma}(x)} \phi_{xx} 
-\frac{\hbar}{\pi} \left( \frac{\partial}{\partial x}\frac{v_{\sigma}(x)}{K_{\sigma}(x)}\right) \phi_x + \frac{B(x)}{\sqrt{2}\pi}, \\
\\
\frac{\partial}{\partial t} \frac{\delta \mathcal{L}}{\delta \phi_t} 
&= \frac{\hbar}{\pi K_{\sigma}(x) v_{\sigma}(x)} \phi_{tt}, \\
\frac{\delta \mathcal{L}}{\delta \phi} &= \frac{\sqrt{2}\pi \hbar v_{\sigma}(x) y_{1\perp}(x)}{(\pi\alpha)^2} \sin(2\sqrt{2} \phi),
\end{align*}
where, according to Sec.~\ref{sec:equation_of_motion}, $\partial_x h(x) = B(x)$ given by Eq.~\eqref{eq:sigmoid}.
Within the bulk of the wire, $v_{\sigma}$, $K_{\sigma}$, $y_{1\perp}$, and $B$ are taken as constants, and we recover the equation of motion
\begin{equation}
\phi_{tt} = v_{\sigma}^2 \phi_{xx} + \frac{\sqrt{2} v_{\sigma}^2 K y_{1\perp}}{\alpha^2} \sin(2\sqrt{2} \phi_{\sigma}) - \frac{B_0 v_{\sigma} K}{\hbar\sqrt{2}},
\label{eq:of_motion}
\end{equation}
where $B_0 = \Delta h/L$.

\section{Critical spin bias}
\label{app:critical_bias}

In the absence of an exact solution to Eq.~\eqref{eq:of_motion} in the wire-and-leads geometry, we find the critical spin bias from the numerical solution. The form of the equation of motion however suggests a simple expression for the critical bias that we use as a reference value. Without the spatial derivative term $v_{\sigma}^2 \phi_{xx}$, Eq.~\eqref{eq:of_motion} describes for instance a Josephson junction driven with external current~\cite{Tinkham1996} or a planar pendulum driven with constant torque. These systems possess a critical drive, given by the coefficient of the cosine, below which a stationary solution exists and above which the solution increases with time.
Based on the similar form of the equation of motion Eq.~\eqref{eq:of_motion}, we expect the transition from insulating ($\partial_t \phi_{\sigma} = 0$) to conducting ($\partial_t \phi_{\sigma} \neq 0$) to occur when the last term of Eq.~\eqref{eq:of_motion} is equal to the coefficient of the sine term. 
We therefore define the expression
\begin{equation}
\Delta h_c = \frac{2 L v_{\sigma} y_{1\perp}}{\alpha^2}
\label{eq:critical_spin_bias}
\end{equation}
for the critical bias and study whether this value is reproduced by the numerical solution.
Here, we again set $\hbar = 1$.

This critical bias can be connected to the threshold voltage $V_{\text{th}}$ in the Landau-Zener dielectric breakdown in band and Mott insulators~\cite{OkaAoki2003, OkaAoki2005, OkaAoki2010, HeidrichMeisnerDagotto2010, TanakaYonemitsu2013, TakasanKawakami2019}. For a one-dimensional insulator of length $L$, numerical results of nonequilibrium Green's function calculations indicate that the dielectric breakdown
is governed by the electric field $E_{\text{th}} = V_{\text{th}}/L \sim \Delta^2/W$, where $\Delta$ is the charge gap and $W$ the bandwidth, when the wire length is larger than the correlation length $\xi = W/\Delta$~\cite{TanakaYonemitsu2013}. For wires shorter than the correlation length, the threshold voltage was found to be approximately given by the gap, $V_{\text{th}} \sim \Delta$~\cite{TanakaYonemitsu2013}. 

We may compare these results to the sine-Gordon model deep in the massive phase ($y_{1\perp} \to \infty$, $K_{\sigma} \to 0$), where an expression for the spin gap is available. The cosine term in the Hamiltonian can be expanded around a minimum, resulting in a quadratic mass term $\sim M^2 \phi_{\sigma}^2$ where the phonon mass
\begin{equation}
M = \sqrt{\frac{4 K_{\sigma} v_{\sigma}^2 y_{1\perp}}{\alpha^2}}
\label{eq:massive_phase}
\end{equation} 
is identified with the spin gap (see Eq.~(2.153) in Ref.~\cite{Giamarchi_one_dimension2003}).
In this limit, we can then expect a threshold magnetic-field gradient 
\begin{equation}
B_{\text{th}} \sim \frac{\Delta_{\sigma}^2}{v_{\sigma}} = \frac{4 K_{\sigma} v_{\sigma} y_{1\perp}}{\alpha^2}.
\end{equation}
This expression is not identical to $\Delta h_c/L$ as given by Eq.~\eqref{eq:critical_spin_bias} but has the same functional dependence on $y_{1\perp}$, $v_{\sigma}$ and the cutoff $\alpha$. In Eq.~\eqref{eq:critical_spin_bias}, the critical bias is independent of $K_{\sigma}$, which agrees with the numerical results in Fig.~\ref{fig:varying_Ksigma}(a) for $K_{\sigma} \geq 0.5$. For $K_{\sigma} < 0.5$, the critical bias is reduced and has a dependence on $K_{\sigma}$. 

To gain insight into this difference, we define the correlation length $\xi_{\sigma} = v_{\sigma}/\Delta_{\sigma}$~\cite{VisuriGiamarchi2020}. Using Eq.~\eqref{eq:massive_phase}, $\xi_{\sigma} = \alpha/(2 \sqrt{K_{\sigma} y_{1\perp}})$ in the strong-coupling limit. Setting $\alpha = 1$ and $y_{1\perp} = 0.5$ as in Fig.~\ref{fig:varying_Ksigma}, we find that the correlation length exceeds the wire length $L = 1$ when $K_{\sigma} < 0.5$. In Ref.~\cite{TanakaYonemitsu2013}, the threshold voltage was found to be approximately equal to the gap in this case. For $K_{\sigma} = 0.2$, shown in Fig.~\ref{fig:varying_Ksigma}(a), the strong-coupling spin gap of Eq.~\eqref{eq:massive_phase} would have the value $\Delta_{\sigma}/\Delta h_0 \approx 0.6$, while the numerically obtained critical bias is $\Delta h_c/\Delta h_0 \approx 0.7$. As the value $y_{1\perp} = 0.5$ is not in the strong-coupling limit, we do not expect the expression~\eqref{eq:massive_phase} for the spin gap to be accurate. Despite this discrepancy, it is interesting to note that the behavior of the critical bias changes when the correlation length exceeds the wire length.

\section{Numerical solution of the equation of motion}
\label{app:discretization}

To solve Eq.~\eqref{eq:equation_of_motion} numerically, we discretize the space and time coordinates as $x_j$ and $t_n$, respectively. We denote the discretized field as $\phi(x, t) \to u_j^n = u(x_j, t_n)$. We reformulate the equation of motion as
\begin{equation*}
\phi_{tt} = v^2 \phi_{xx} + \delta_{\sigma} \phi_x + F(\phi(x, t), x, t),
\end{equation*}
where
\begin{equation}
\delta_{\sigma} = \frac{v_{\sigma}(x)}{K_{\sigma}(x)} \left[ v_{\sigma}'(x) K_{\sigma}(x) - v_{\sigma}(x) K_{\sigma}'(x) \right].
\end{equation}
In the discretized form, this is
\begin{align*}
\frac{u_j^{n + 1} - 2u_j^n + u_j^{n - 1}}{\Delta t^2} = &v_j^2 \frac{u_{j + 1}^n - 2 u_j^n + u_{j - 1}^n}{\Delta x^2} \\
&+ \frac{v_j^2 \delta_{j}}{2 \Delta x} \cdot \frac{u_{j + 1}^n - u_{j - 1}^n}{2 \Delta x} \\
&+ F(u_j^n, x_j, t_n) + \mathcal{O}(\Delta x^2, \Delta t^2),
\end{align*}
where
\begin{equation*}
\delta_{j} = \left( \frac{v_{j + 1} - v_{j - 1}}{v_j} - \frac{K_{j + 1} - K_{j - 1}}{K_j} \right)
\end{equation*}
and
\begin{align*}
F(u_j^n, x_j, t_n) = - \frac{\sqrt{2} v_j^2 K_j y_j}{\alpha^2} \sin(2\sqrt{2} u_j^n) + \frac{B_j v_j K_j}{\sqrt{2}}.
\end{align*}
We may now rewrite the equation as
\begin{align*}
u_j^{n + 1} = - u_j^{n - 1} + &2(1 - \gamma_j^2) u_j^n + \gamma_j^2 \left( u_{j + 1}^n + u_{j - 1}^n \right) \\
&+ \frac{\delta_j}{4} \gamma_j^2 (u_{j + 1}^n - u_{j - 1}^n) \\
&+ \Delta t^2 F(u_j^n, x_j, t_n) + \mathcal{O}(\Delta x^2, \Delta t^4),
\end{align*}
where $\gamma_j = v_j \frac{\Delta t}{\Delta x}$ and now $j \in \{2, \dots, J - 1\}$ and $n \in \{2, \dots, N - 1\}$. 

The non-reflecting (Sommerfeld) boundary conditions of Eq.~\eqref{eq:boundary_conditions} are discretized as
\begin{align*}
u_J^{n + 1} &= u_{J - 1}^n - Q_J u_{J - 1}^{n + 1} + Q_J u_J^n, \\
u_1^{n + 1} &= u_2^n - Q_1 u_2^{n + 1} + Q_1 u_1^n,
\end{align*}
where $Q_j = (1 - \gamma_j)/(1 + \gamma_j)$. As initial conditions, we set $u_j^1 = u_j^2 = 0$.
We set the discretization as $\Delta x/L = 5 \times 10^{-5}$ and $\Delta t/(L/v_F) = 2 \times 10^{-5}$ while the width $w$ of the wire-lead boundary in Eq.~\eqref{eq:sigmoid} is set to $w/L = 10^{-2}$.

\section{Initial time evolution in a Luttinger liquid wire}
\label{app:LL_wire}

In the case of an interacting Luttinger liquid wire, the equation of motion is~\cite{Maslov_conductance1995}
\begin{equation}
\phi_{tt} = v^2 \phi_{xx} + K v \phi_x \partial_x \left(\frac{v}{K}  \right) + \sqrt{2}K v E.
\label{eq:equation_of_motion}
\end{equation}
Here, $E$ is the electric field within the wire due to a chemical potential bias which drives charge transport. In analogy with the spin transport, we assume a field $E$ of the form of Eq.~\eqref{eq:sigmoid}.
A model of this form arises, in one dimension, for spinless fermions or the charge sector of fermions with spin. The parameters $v$ and $K$ are here the ones pertaining to the charge sector.

\begin{figure}
\centering
\includegraphics[width=\linewidth]{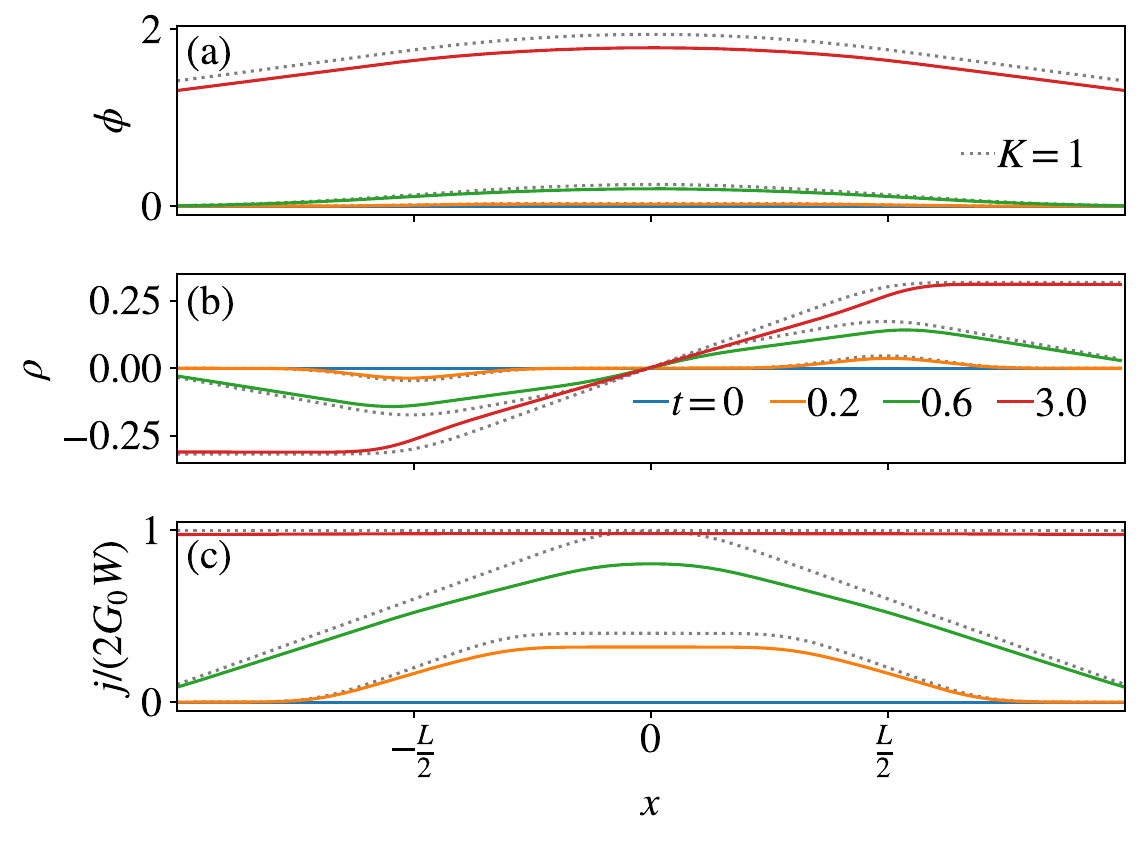}
\caption{The spatial profiles of (a) the field $\phi$, (b) the charge density $\rho = -\frac{\sqrt{2}}{\pi} \partial_x \phi$, and (c) the current $j = \frac{\sqrt{2}}{\pi} \partial_t \phi$, at fixed times during the initial time evolution. The solid lines correspond to $K = 0.8$ in the wire and $K_L = 1$ in the leads, while the dotted lines show the noninteracting case $K = 1$ at the same times for comparison. Time is in units of $L/v_F$, and we set $v = v_F$ everywhere and the bias to $V/W = 0.5$. Here, $W$ is the energy scale $W = 1/(\pi \rho_0)$ with the constant density of states $\rho_0 = 1/(\pi v_F)$.}
\label{fig:profiles_LL}
\end{figure}

As a comparison to Sec.~\ref{sec:initial}, we present here the initial time evolution of $\phi$, the charge density $\rho = -\frac{\sqrt{2}}{\pi} \partial_x \phi$, and current $j = \frac{\sqrt{2}}{\pi} \partial_t \phi$. These are shown in Fig.~\ref{fig:profiles_LL} for both an interacting Luttinger liquid wire coupled to noninteracting leads and a fully noninteracting system. The electric field within the wire is quenched to a nonzero value at time $t = 0$, which induces a charge current between the contacts. Charge density is depleted from the left reservoir and accumulates in the right one, similar to spin density in the case of spin transport. 
As the wire is ballistic, there is no resistance to charge transport and the current profile evolves into $j = 2G_0 V$ everywhere in the wire, as seen in Fig.~\ref{fig:profiles_LL}(c). Here, $V$ is the voltage, or chemical potential bias, $V = E_0 L$. The charge current has a factor of $2$ due to spin degeneracy. The steady-state current is independent of $K$ in the wire~\cite{Safi_transport1995, Maslov_conductance1995, Ponomarenko_conductance1995}.

\section{Parameters in the cold-atom experiment}
\label{app:experimental}

Similar to Ref.~\cite{VisuriGiamarchi2020}, we determine the parameters $y_{1\perp}$, $K_{\sigma}$, and $v_{\sigma}$ from the experimental parameters used in Ref.~\cite{Krinner_spin_and_particle2016}. We consider a one-dimensional channel, corresponding to a situation where only one transverse mode is occupied. 
The effective parameters $y_{1\perp}$ and $K_{\sigma}$ can be determined from the experimental parameters via the Gaudin-Yang model of Eq.~\eqref{eq:continuum_hamiltonian},
which describes spin-$\frac{1}{2}$ fermions with interacting with an effective delta-function interaction potential. 
When the atoms are confined into a quasi-one-dimensional geometry, the coupling $g_{1\perp}$ is obtained as 
\begin{equation}
g_{1\perp} = \frac{2\hbar \omega_{\perp}a}{1 - A \frac{a}{a_{\perp}}},
\end{equation}
where $\omega_{\perp}$ is the transversal confinement frequency, $a$ the $s$-wave scattering length, $a_{\perp}$ the oscillator length, $m$ the atom mass, and $A$ a dimensionless constant arising from the wave-function expansion in the derivation of the effective interaction potential~\cite{Olshanii_atomic_scattering1998}. The oscillator length is given by $a_{\perp} = \sqrt{\hbar/(m\omega_{\perp})}$. The values used for these parameters are given in Table~II of Ref.~\cite{VisuriGiamarchi2020}. 

The dimensionless coupling $y_{1\perp}$ is obtained from $g_{1\perp}$ as
\begin{equation}
y_{1\perp} = \frac{g_{1\perp}}{\pi v_{\sigma}}.
\end{equation}
The oscillation behavior of the spin current is independent of the sign of $y_{1\perp}$ as it only produces a phase shift in $\cos(2\sqrt{2} \phi_{\sigma})$. For spin-rotation invariant interactions, $K_{\sigma}$ is fixed by
\begin{equation}
K_{\sigma} = \sqrt{\frac{1 + \frac{y_{1\perp}}{2}}{1 - \frac{y_{1\perp}}{2}}}
\end{equation}
(see for example Eq.~(2.105) in Ref.~\cite{Giamarchi_one_dimension2003}). The spin velocity is obtained by interpolating between the analytic expressions for weak and strong attractive interactions~\cite{Fuchs_exactly_solvable2004}
\begin{align}
\frac{v_{\sigma}}{v_F} \simeq 
\begin{cases}
1 - \frac{\gamma}{\pi^2} + \dots, &\frac{1}{\gamma} \to -\infty, \\
-\frac{\gamma}{\pi\sqrt{2}} \left(1 - \frac{2}{\gamma} + \dots \right), &\frac{1}{\gamma} \to 0^{-},
\end{cases}
\label{eq:spin_velocity}
\end{align} 
where $\gamma = m g_{1\perp}/(\hbar^2 \rho_0)$. In the noninteracting system with $K_{\sigma} = 1$, the spin velocity is equal to the Fermi velocity $v_F = \hbar \rho_0 \pi/(2 m)$. We choose the short-distance cutoff $\alpha$ as the inverse density $\rho_0^{-1}$. 

We similarly compute the exact spin gap~\cite{Fuchs_exactly_solvable2004}:
\begin{align}
\frac{\Delta_{\sigma}}{E_F} \simeq 
\begin{cases}
\frac{16}{\pi} \sqrt{\frac{|\gamma|}{\pi}} e^{-\pi^2/(2|\gamma|)} + \dots, 
&\frac{1}{\gamma} \to -\infty, \\
\frac{2\gamma^2}{\pi^2} \left[ 1 - \frac{\pi^2}{2\gamma^2} + \mathcal{O}\left(\gamma^{-4}\right)\right],  &\frac{1}{\gamma} \to 0^{-},
\end{cases}
\label{eq:exact_spin_gap}
\end{align}
where $E_F = m v_F^2/2$ is the Fermi energy. 
Using Eqs.~\eqref{eq:spin_velocity} and~\eqref{eq:exact_spin_gap}, an order-of-magnitude estimate for the critical bias can be obtained as $\Delta h_{\text{LZ}} = L \Delta_{\sigma}^2/v_{\sigma}$. The values of the spin velocity, spin gap, correlation length, and the critical bias are computed in Table~\ref{tab:critical_bias} for the particle density $\rho_0 = 0.8/\mu$m and wire length $L = 6 \mu$m. The scattering lengths are given in units of the Bohr radius $a_0 = 5.29 \times 10^{-11}$ m, the Boltzmann constant is $k_B = 1.38 \times 10^{-23}$ J$/$K, and the reduced Planck constant $\hbar = 1.054 \times 10^{-34}$ Js. As the correlation length $\xi_{\sigma}$ is smaller than the wire length but comparable to it, it is not unambiguous whether $\Delta h_{\text{LZ}}$ or $\Delta_{\sigma}$ is a better estimate for the critical bias (see Appendix~\ref{app:critical_bias}). This remains to be determined experimentally.

\begin{table}[h]
\caption{The spin velocity and spin gap given by Eqs.~\eqref{eq:spin_velocity} and~\eqref{eq:exact_spin_gap} for $\rho_0 = 0.8/\mu$m and $L = 6 \mu$m. These are used to estimate the correlation length $\xi_{\sigma} = \hbar v_{\sigma}/\Delta_{\sigma}$ and the critical spin bias $\Delta h_{\text{LZ}} = L \Delta_{\sigma}^2/(\hbar k_B v_{\sigma})$ due to a Landau-Zener-like transition.}
\begin{tabular}{c | c | c | c | c}
$a/a_0$ &$v_{\sigma}$ (mm$/$s) &$\Delta_{\sigma}/k_B$ (nK) &$\xi_{\sigma}$ ($\mu$m)	&$\Delta h_{\text{LZ}}$ (nK) \\
 \hline
$-2623$ &16  &25  &4.8	&32  \\
$-3789$ &17  &62  &2.1	&180 
\end{tabular}
\label{tab:critical_bias}
\end{table}

\bibliographystyle{apsrev4-1-with-titles}
\bibliography{bibfile}

\end{document}